# Studying Diffusion of Viral Content at Dyadic Level

Anita Zbieg[1], Błażej Żak[2], Jarosław Jankowski[3], Radosław Michalski[2], Sylwia Ciuberek[2]

[1] Institute of Psychology, University of Wrocław, Wrocław, Poland
[2] Faculty of Computer Science and Mgmt., Wrocław University of Technology, Wrocław, Poland
[3] Faculty of Computer Science, West Pomeranian University of Technology, Szczecin, Poland
anita.zbieg@gmail.com, jjankowski@wi.zut.edu.pl, {blazej.zak, radoslaw.michalski, sylwia.ciuberek}@pwr.wroc.pl

*Abstract*—Diffusion of information and viral content, social contagion and influence are still topics of broad evaluation. As theory explaining the role of influentials moves slightly to reduce their importance in the propagation of viral content, authors of the following paper have studied the information epidemic in a social networking platform in order to confirm recent theoretical findings in this area. While most of related experiments focus on the level of individuals, the elementary entities of the following analysis are dyads. The authors study behavioral motifs that are possible to observe at the dyadic level. The study shows significant differences between dyads that are more vs less engaged in the diffusion process. Dyads that fuel the diffusion proccess are characterized by stronger relationships (higher activity, more common friends), more active and networked receiving party (higher centrality measures), and higher authority centrality of person sending a viral message.

## I. INTRODUCTION

Researchers from various fields, studying diffusion of innovation process [1], [2], [3], social influence mechanism [4], [5], [6], social contagion and epidemics outbreaks [7], [8], cascades of influence patterns [9], word of mouth process [10], [11] or viral marketing seeding strategies [12], [13] are investigating a common phenomenon: a diffusion of information (content, opinions, behaviors or emotions) within a network of social relations. Fundamentally the mechanism of social propagation can be explained in two folds: by *influence* when individuals intentionally and directly influence others, or by *imitation* when they become imitated by them [14]. However, more detailed investigations related to the process itself and its important factors are rather ambiguous. Two-step flow model of communications [15], considered a small group of people called *influentials* as important for social influence and diffusion process, as they directly influence many neighbors. Thus, influential opinion leaders and their characteristics attract social contagion researchers from decades and are still popular in theory [16], empirical studies [17] and business practices [18], [19]. However, some more recent research brought new findings and questions. Computer simulations and empirical data showed that hypothetical *influentials* might be rare and difficult to observe in a social influence process [20]. Global social changes are more often driven by easily influenced individuals, influencing other easily influenced individuals [20]. Cascades of influence are mostly small, occur in first degree from sender, and rarely derive from chains of referrals [21]. Finally, well connected individuals having many neighbors, do not necessarily have bigger influence than others [22]. Maintaining many connections can be costly, and that can be a reason why individuals with many ties more often form weaker relationships by contrast to individuals with fewer but stronger ties [23]. Viral content distributed by strongly connected individuals might be less relevant to receiver preferences, interests or personality and more often ignored [22]. Attracted by those discrepancies, authors of the following paper analyzed viral content propagation in a communication network observed within a virtual world avatar chat website.

## II. RELATED WORK

Empirical research related to the question of factors driving social contagion come from diverse studies and generally three groups of factors are considered important: message characteristics, individual sender or receiver characteristics and social network characteristics. What showed YouTube videos diffusion study [22], in viral content propagation, number of seeds is an important factor determining the reach of a viral campaign. The more initial seeds company is able to reach, the larger number of users will become exposed to the message, but the quality of viral content and homogeneity of the influenced group also play a significant role in the process. Another important characteristic is the number of connections a seed individual has in a network. This measure, often used for determining importance of an individual, has proven to have negative effect on the size of an information cascade within a viral videos diffusion study [22]. The more connections seeding individual had, the smaller information cascade has been observed, and viral content received from individuals with many ties had bigger chance to be ignored. Similarly, in a communication patterns study performed on data gathered from Twitter, Facebook and Yahoo social networks [21], cascades of information diffusion were more likely to occur between people with low to moderate number of connections. People with many connections seem to spread viral content less effectively (than those with few ties) and are more difficult to activate because they receive information from many other sources [20]. Important for the process is also an affinity of preferences between connected individuals. In other words, the more homogenous preferences are in a group of seed's neighbors, the more viral adoptions are observed. Another empirical viral marketing study conducted on social network sites and in a mobile network [13] showed that seeding strategies are the most effective if aimed at well-connected individuals: hubs with many connections, or bridges that connect different groups. What is interesting is that, well connected individuals seem to be better seeds not because they are more persuasive, but because they are more active and more willing to participate in a diffusion process. Additionally, as they participate in a network more frequently, they are more often exposed to anything that flows through the network including a virus or a viral content [8], [13]. The study of viral spread of avatar gestures in a multiplayer virtual world of Second Life [24] showed that adoption rules in this network depend mainly on the number of friends adopting. Individuals with many friends were less likely to be influenced by others and this is what researchers explain by weaker tie strength formed by highly connected individuals. Important for the process are also characteristics of the asset (popular, niche) spreading through the network, and a strength of tie that can be measured by the number of common friends or triples formed by two individuals [26].

The above research findings about the roles and factors important for diffusion process are sometimes ambiguous or contradictory and that is what motivated authors to study the topic. Relationships are foundations of a network hence it seems important to study them in depth. Communication events in online social networks are the moments when relationships become visible. Studying dyads is a way to observe those communication events on the very basic level. The complexity of interactions that can occur at the dyadic level (between two people) in real world seems unlimited and well-studied by social psychologists. However virtual environments strictly define the list of possible interactions that can occur within a dyad, there is still a large room for research in this area.

III. THEORETICAL FRAMEWORK

When analyzing social networks, a researcher can focus on five distinct levels of analysis: individual actor, dyadic, triadic, subgroup or global level [25]. In this paper, the basic unit of analysis are dyads that are composed of two individuals and having a linkage between them. On the dyadic level, the authors reflect on the relation observed via interactions between two people that can be considered as a channel for spreading information [27], attitudes, emotions or behaviors [28]. Three basic elements necessary for communication to occur: a sender, receiver and communication channel [27] are present in every dyad, and this makes this perspective attractive for viral information diffusion study. Using the construct of dyad it is possible to reflect and characterize this basic communication situation or event as entity. In other words, dyads are considered by authors of the paper as basic units of observations, in which communication happens. Dyads are defined with sender, channel (relation) and receiver characteristics (see Figure 1). Sender and receiver are understood as a role played in the situation of communication event and the relation is a channel through which a message flows.

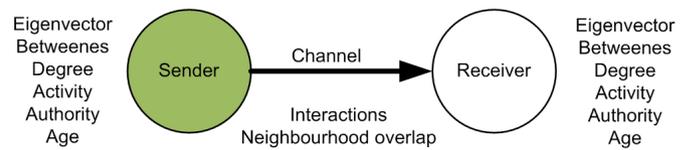

Figure 1. Dyad as a communication event

To characterize dyads, attributes of its three basic elements (sender, channel and receiver) have been calculated. The relationship measures that have been investigated are the number of messages exchanged and the number of common friends shared by the sending and receiving user. In this paper, the information about the communication activity, network position, and basic demographics (age, gender) for the sender and receiver (influencer and influenced) has been investigated. To describe network position of both users forming a dyad commonly used measures have been calculated: eigenvector, authority and betweenness centralities, degree and overall activity. The centrality metrics represent different aspects of the user's network position: authority centrality represents a leadership in strong cliques or the importance of a person in the means of information acquiring possibilities; betweenness centrality shows an ability of being an information broker [25]; and eigenvector centrality is a common measure of influence of a user in a network.

In this study an influencer is defined as a person that *transmits* viral message that influences other person directly (not by imitation). Influenced is a person who *receives* a message and is an object of influence. This person can *transmit* the viral message further, but can also *use* the information (e.g. watch video, buy a product, go to an event etc.). Hence, there are two behaviors distinguished and defined: *transmission* - sending and receiving a viral message, and *infection* - that means making use of it. The person influenced can behave differently when exposed to a viral message. There are four common behavior patterns among the dyads that create an information cascade as presented in Figure 2.

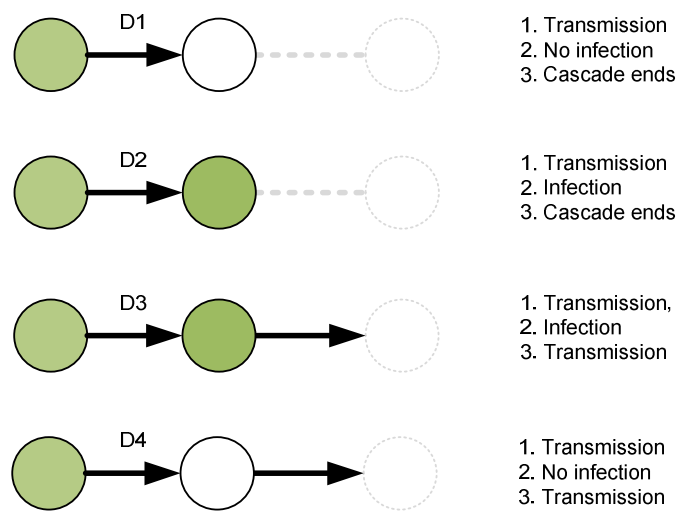

Figure 2. Behavior patterns investigated

Dyad groups can be briefly described as:

- D1 is a dyad in which the influencer sent a message but did not influence the receiver neither to send it further or use it,
- D2 is a dyad in which the influencer succeeded into influencing the receiver to use the information,
- D3 is a dyad that propagated cascade further, where both users used and transmitted the information,
- D4 is a dyad where the person that received the viral information did not use it but only transmitted it further.

In the study, the authors consider communication situations in which a viral message is propagated. The research sample is composed with dyads through which the message has been transferred. The observed communication events (dyads) are divided into groups (Figure 2) that are characterized by different behaviors of the receiving party (use a message or not, transfer a message or not).

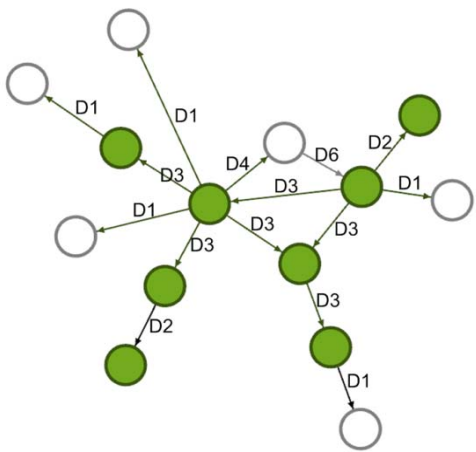

Figure 3. Dyadic motifs in the network

To illustrate the intuition behind the proposal of the work let's consider a small cascade illustrated above (see Figure 3). Each relation on the graph represents one communication event that involved transmission of the virus. By observing dyadic motifs it is possible to label relations according to role played in the cascade propagation. As dyad is a basic building block of an information cascade it is worth studying what properties may differentiate different kinds of dyads. This leads to more detailed questions. How a dyad ending a cascade (D1) differs from a dyad that fuels the diffusion process (D3)? Which communication events are more likely to lead to usage of information propagated with a virus? Is it possible to tell which dyads are more likely to take part in an information cascade spreading through the network?

The aim of the study was to find out, which of the characteristics of communication situations are important for infections and further viral message propagation. By comparing the groups of dyads, the authors tried to find out which part of a dyad: sender, relation or receiver; and which of its characteristics (activity, centrality, tie strength) significantly differentiate the behavior patterns important for a viral message diffusion process. The following hypothesis can be proposed for an investigation:

H1: Influencers that successfully persuade others to transfer a viral message further are characterized by higher eigenvector centrality than others.

H2: People that are active will transfer the viral message more often than others after receiving it.

H3: Viral message transferred through strong relationship is more infectious than message transferred through weak channel. Viral message transferred through strong ties is also more often transmitted further.

IV. STUDY AND DATA ANALYSIS

A. Social Network Description

The study covers a time of two 5 day periods of March 2012. During the first period (T1), a data about public chat communication was recorded and analyzed to compute the characteristics of each dyad in the network. During the second period (T2) the viral diffusion was observed. In T2 two types of behaviors were distinguished: propagation of the viral message (transfer) and usage of the viral message (infection). A specific limitation of the virtual world mechanics constraints was that it was impossible to imitate other users without interacting with them. To get the item, the user had to receive it from another person and this made the diffusion observable step by step. The viral content spreading through the network was a Guy Fawkes "Anonymous" mask that users could put on their avatars to express "against ACTA" opinion in a virtual protest that took place in the network. A person was considered *infected* when he/she put on the mask on his/her own avatar. The user could also *transfer* the mask by sending the viral message further.

The authors investigated a network built of messages exchanged between users in the period T1. General properties of the network are presented in Table I.

TABLE I. ANALYZED SOCIAL NETWORK DESCRIPTION

| Property | Value |
|---|---:|
| Number of nodes | 2362 |
| Number of edges | 25134 |
| Average node degree | 10.64 |
| Network diameter | 9 |
| Average path length | 3.233 |

Figure 4 shows the whole social network captured at T1 (grey links) together with the map of viral message diffusion captured at T2 (red links and nodes). The anonymous mask (viral content) was initially delivered to 16 users who simply asked for it and were the most involved in the virtual protest. They have seeds for the viral transmissions, and the only way to obtain the object later on was receiving it from other network participants. Among 4910 users that logged in during

T2, 324 of them (6.60%) received viral content and 134 of them decided to send it to a friend. That made 41% of receivers to be engaged in forwarding the message.

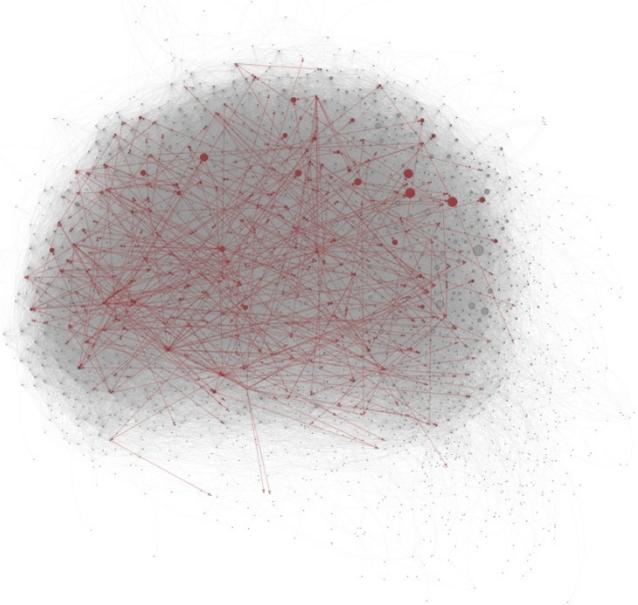

Figure 4. Visualization of the analyzed social network (edges marked red - viral message diffusion paths, node size - user activity)

TABLE II. DYAD GROUPS OVERVIEW

| Group | Group size | Average neighborhood overlap | Average number of interactions |
|---|---|---|---|
| D1 | 89 | 32.32 | 32.45 |
| D2 | 156 | 37.02 | 42.59 |
| D3 | 146 | 52.16 | 33.09 |
| D4 | 22 | 48.77 | 22.40 |

The main units of analysis within the study are dyads, grouped according to behavior pattern they have expressed within the viral message diffusion process (T2). The size and properties of extracted groups are presented in Table II, where the neighborhood overlap is a number of common neighbors in the network and the number of interactions means the number of messages exchanged between dyad members in the analyzed period (T1).

B. Behavior Comparison: Infection & Message Transfer

After having dyads assigned to groups and their characteristics calculated, the first test was performed. No significant differences has been found between dyads that led (or not) to the infection of the receiving party (D1 ∪ D4) vs. (D2 ∪ D3). Tests were performed by using Mann-Whitney U test [29] in order to calculate differences in all the observed properties between dyads (D1 ∪ D4) and (D2 ∪ D3). Mann-Whitney U test was chosen because it does not expect the distributions of the observed variables to follow the normal distribution. Distributions of a user activity within a network and centrality measured more often follow a power law than fit the Gauss curve. This was the main reason of choosing this statistical procedure.

On the other hand there are a lot of differences between communication events that led (or not) to further propagation of the viral content (D1 ∪ D2) vs. (D3 ∪ D4). Dyads that engaged the receiver to pass the virus further have been characterized by stronger relationship (higher activity, more common friends) and more active and networked receiving party (higher centrality measures). There were also significant differences between authority centrality of an influencer in the dyads (D1 ∪ D2) vs. (D3 ∪ D4). In other words senders with higher authority centrality more often did convince the receiver to pass the content further. All significant differences have been summarized in Table III. This analysis lets us to reject the first hypothesis (H1). The influencer eigenvector did not differentiate dyads that has led to further propagation of the viral content (D3 ∪ D4). More useful measure of influence in this study was influencer authority centrality that significantly differentiates dyads where the sender influenced the receiver to transfer the message further (see Table III).

Test results also confirm the hypothesis H2. People who had been more active during T1 (influenced the activity measure) and those that received the viral message during T2, were more likely to transfer it further (dyads D3 ∪ D4).

TABLE III. MANN-WHITNEY U TEST RESULS FOR GROUP D1∪D2 VS. D3∪D4 FOR P < 0.05

| | Mean rank D1∪D2 | Mean rank D3∪D4 | p-value | Valid N D1∪D2 | Valid N D3∪D4 |
|---|---|---|---|---|---|
| Influencer authority | 196.09 | 222.91 | 0.025 * | 245 | 168 |
| Relation interactions | 196.26 | 222.66 | 0.027 * | 245 | 168 |
| Relation neighborhood overlap | 184.44 | 239.90 | 0.000 ** | 245 | 168 |
| Influenced activity | 187.11 | 236.01 | 0.000 ** | 245 | 168 |
| Influenced degree | 186.63 | 236.70 | 0.000 ** | 245 | 168 |
| Influenced betweenness | 187.08 | 236.05 | 0.000 ** | 245 | 168 |
| Influenced eigenvector | 185.68 | 238.10 | 0.000 ** | 245 | 168 |
| Influenced reputation | 185.51 | 238.34 | 0.000 ** | 245 | 168 |

*significance level 0.05, ** significance level 0.01

## C. Dyad Comparison Procedure

To find out attributes that differentiate behavior motifs represented by dyads D1...D4 a similar approach to the above has been used. Each group of observations (D1...D4) has been compared against each other to find out significant differences in dyad characteristics:

- Group D1 was compared against D2∪D3∪D4;
- Group D2 was compared against D1∪D3∪D4;
- Group D3 was compared against D1∪D2∪D4;
- Group D4 was compared against D1∪D2∪D3;

By comparing dyads in this manner it is possible to find attributes characteristic for a given dyadic motif, that are significantly lower or higher than in other groups. Because performing all the comparisons involved testing 72 hypotheses using Mann-Whitney U test, the results have been summarized symbolically in tables IV, V and VI. If there was a significant difference found in a particular comparison, the arrow direction shows whether values were higher or lower than in other groups, while "-" shows that no significant differences have been found.

## D. Evaluating Differences Between Sender Characteristics within Investigated Dyads

Table IV presents sender's characteristics that differentiate dyads D1...D4 across the investigated attributes: activity, reputation, degree, age and centrality metrics. No significant differences have been found among those characteristics in dyads D1….D3. This means that none of the investigated measures differentiate influencers across those behavior motifs.

On the other hand influencers in dyad D4 (where recipient did not use the information but only transferred it further), had significantly higher degree, eigenvector and authority centrality than influencers in other dyads (D1…D3). However dyad D4 was represented only by 22 observations and hence no farfetched conclusions from this result should be drawn.

## E. Evaluating Differences Between Receiver Characteristics within Investigated Dyads

The same procedure has been applied to investigate differences between the receiver characteristics across the investigated dyads (Table V). A significantly higher activity and centrality measure has been observed in dyad D3. This means that more active and central users (during T1) were more likely to get infected and pass the message further after they had received it (during T2).

On the other hand dyads where the receiver got infected but did not transfer the message further (D2) has had less active and less central receiving party. In other words users occupying peripheries of a network can be more easily influenced than others (in terms of using the information), but they are less likely to pass the message further.

## F. Evaluating Differences Between Relationship Characteristics within Investigated Dyads

Dyad relationships has been analyzed according to the number of interactions, influencer and influenced activity (number of messages exchanged) and neighborhood overlap (number of common friends). The results of the test are shown in Table VI. Dyad D3 contained significantly stronger relationships (more common friends) with higher activity (during T1). This means that receiving users were more likely to get infected and pass the message further when they have received it from a person who share a lot of common friends and has been exchanging more messages during the period preceding the diffusion (T1). This result confirms the third hypothesis (H3).

On the other hand, the message received from a person that shared a few common friends was less likely to be transferred further (dyad D2).

TABLE IV. INFLUENCER CHARACTERISTICS THAT DIFFENTIATE DYADS D1..D4

| | n | Activity | Reputation | Degree | Betweenness | Eigenvector | Authority | Age |
|---|---|---|---|---|---|---|---|---|
| D1 vs. (D2 ∪ D3 ∪ D4) | 89 | - | - | - | - | - | - | - |
| D2 vs. (D1 ∪ D3 ∪ D4) | 156 | - | - | - | - | - | - | - |
| D3 vs. (D1 ∪ D2 ∪ D4) | 146 | - | - | - | - | - | - | - |
| D4 vs. (D1 ∪ D2 ∪ D3) | 22 | - | - | ↑ * | - | ↑ * | ↑ * | - |

TABLE V. INFLUENCED CHARACTERISTICS THAT DIFFENTIATE DYADS D1..D4

| | n | Activity | Reputation | Degree | Betweenness | Eigenvector | Authority | Age |
|---|---|---|---|---|---|---|---|---|
| D1 vs. (D2 ∪ D3 ∪ D4) | 89 | - | - | - | - | - | - | - |
| D2 vs. (D1 ∪ D3 ∪ D4) | 156 | ↓ ** | ↓ ** | ↓ ** | ↓ ** | ↓ ** | - | - |
| D3 vs. (D1 ∪ D2 ∪ D4) | 146 | ↑ ** | ↑ ** | ↑ ** | ↑ ** | ↑ ** | ↑ ** | ↓ * |
| D4 vs. (D1 ∪ D2 ∪ D3) | 22 | - | ↑ * | - | - | - | - | - |

*significance level 0.05, ** significance level 0.01*

TABLE VI. RELATION CHARACTERISTICS THAT DIFFENTIATE DYADS D1..D4

|  | n | Interactions | Influencer activity | Influenced activity | Neighborhood overlap |
|---|---|---|---|---|---|
| D1 vs. (D2 ∪ D3 ∪ D4) | 89 | - | - | - | - |
| D2 vs. (D1 ∪ D3 ∪ D4) | 156 | - | - | - | ↓ ** |
| D3 vs. (D1 ∪ D2 ∪ D4) | 146 | ↑ * | ↑ * | ↑ * | ↑ ** |
| D4 vs. (D1 ∪ D2 ∪ D3) | 22 | - | - | - | - |

*significance level 0.05, ** significance level 0.01*

## V. CONCLUSIONS

It should be noticed that no significant differences has been found between communication events (dyads) that led (or not) to infection of the receiving party (D1 ∪ D4) vs. (D2 ∪ D3). Using the information, like joining a protest, or buying a product seems to be independent from relationship, network or activity measures that have been observed. Perhaps it depends mostly on individual preferences of the receiving party that have not been captured during the study. On the other hand there are a lot of differences between communication events that led (or not) to further propagation of the viral content (D1 ∪ D2) vs. (D3 ∪ D4). Dyads that engaged the receiver to pass the message further have been characterized by stronger relationship (higher activity, more common friends), more active and networked receiving party (higher centrality measures), and higher authority centrality of influencer.

This short analysis shows that different types of behaviors may depend completely on different factors. An activity of passing a viral content has shown to be dependent on network centrality measures of both users involved in the communication process. Contrary - the activity of putting the mask on (in other words using the information), has shown to be not dependent on any of the observed variables. This might be one of the reasons why different conclusions from different studies may be encountered in literature. Propagating the information further might not be exactly the same behavior among different social networks. And certainly buying a product is not the same as tweeting about it. The following study shows that a behavior of sharing the information in many ways depends on a network we are embedded in, and information received from *authority* is more likely to be shared further. However the decision about using it (at least in this study) is individual and independent from the network. That is understandable when social media are considered as an information distribution channel. The way information travels is surely dependent on a network underneath, however decision on who will use the information and in what way, is a consequence of an individual decision of each person involved in the viral process.

This shows the limits of influence an authority or influential can have on a network around. Being active and important part of a network those people able to distribute the message more broadly than an average network participant, however if the message does not satisfy individual preferences of people who receive it, it will most likely get ignored. The study shows that if one would like to effectively distribute viral contents (dyad D3) he or she should send it to close friends (active relationship and many common friends) that are the most active and central in the network. These kinds of ties seem to be most likely to trigger infection and further propagation of the content by the receiving party.

## VI. FUTURE WORK

However dyads that took an active part in the information diffusion process (D1...D4) seem to be most important in the study, one should be aware that there are other possible roles to be considered at the dyadic level. To have a full picture of what has happened on the dyadic level one might investigate other possibilities (see Figure 5). An interesting group of behavior motifs that unfortunately has been underrepresented in this study (only 22 cases) are dyads D5...D8 that exemplify behavior of information brokerage (when sender only passes a message further without actually using the information). From a viral marketing perspective it is vital to distinguish users that share information about a product from those that actually have bought it. Although both groups are important, they play different roles in product promotion and adoption process, and perhaps they also differ in their characteristics. As it has been already pointed out, there are specific dyad characteristics that make it more likely to fuel a viral information diffusion process. It might be interesting to find out characteristics of dyads that are stopping it. Dyads D9...D16 represent boundaries of a diffusion process. The question of what differentiate dyads that stopped the viral message is still open for an investigation.

Network motif analysis is a useful tool for finding out patterns characteristic of a given network [30], sometimes called "network fingerprints" [31]. The same technique can be applied to characterize and compare viral messages at the dyadic level. If one can track a message and capture a social network through which the message is spreading, he is able to calculate the number of motifs (D1...D16) and to create a message profile telling us which motif is over or under-represented in the diffusion process. Such information should be valuable for comparing different viral campaigns, assessing their strengths and weaknesses, and finding out which motifs play the most important role in a message diffusion. Also, having an insight into the characteristics of dyads representing each motif should be valuable for understanding how the process of information diffusion depend on a network it is embedded in.

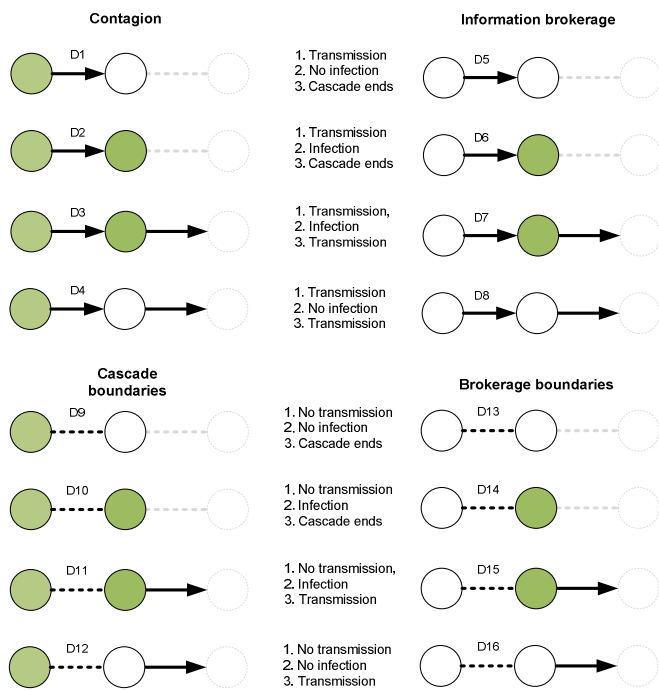

Figure 5. Behavior motifs to be analyzed at the dyaic level

ACKNOWLEDGMENT

The work was partially supported by fellowship co-financed by the European Union within the European Social Fund, the Polish Ministry of Science and Higher Education, the research project 2010-13. Calculations have been carried out in Wroclaw Centre for Networking and Supercomputing (http://www.wcss.wroc.pl), grant No 177.